\documentclass[
reprint,
superscriptaddress,
altaffilletter,
nofootinbib,
amsmath,
amssymb,
aps,
prd]{revtex4-1}

\usepackage{times}
\usepackage{mathtools}
\usepackage{graphicx}
\usepackage{dcolumn}
\newcolumntype{d}[1]{D{.}{.}{#1}}
\usepackage{bm}
\usepackage{soul}
\usepackage{verbatim} 
\usepackage{cprotect} 
\usepackage{url}
\usepackage{subfigure} 
\usepackage{tikz}
\usepackage{afterpage}
\usepackage{nicefrac}
\usepackage{tabularx}
\newcolumntype{L}{>{\raggedright\arraybackslash}X}
\usepackage{multirow}
\usepackage[english]{babel}
\usepackage{natbib} 
\usepackage{glossaries} 
\usepackage{hyperref}
\usepackage{aas_macros} 
\usepackage{comment}
\usepackage{float}
\usepackage{placeins}
\usepackage{ulem}
\graphicspath{{figures/}}

\usepackage{xcolor} 

\begin{document}

\title{Detectability of lensed gravitational waves in matched-filtering searches}
\author{Juno C. L. \surname{Chan}}
\email{chun.lung.chan@nbi.ku.dk}
\affiliation{Center of Gravity, Niels Bohr Institute, Blegdamsvej 17, 2100 Copenhagen, Denmark}

\author{Eungwang \surname{Seo}}
\affiliation{SUPA, School of Physics and Astronomy, University of Glasgow, Glasgow G12 8QQ, United Kingdom}

\author{Alvin K. Y. \surname{Li}}
\affiliation{RESCEU, The University of Tokyo, Tokyo, 113-0033, Japan}
\affiliation{Department of Physics, The Chinese University of Hong Kong, Hong Kong}
\affiliation{LIGO Laboratory, California Institute of Technology, Pasadena, California 91125, USA}

\author{Heather \surname{Fong}}
\affiliation{University of British Columbia, Vancouver Campus, 2329 West Mall, Vancouver, British Columbia V6T 1Z4, Canada}

\author{Jose M. \surname{Ezquiaga}}
\affiliation{Center of Gravity, Niels Bohr Institute, Blegdamsvej 17, 2100 Copenhagen, Denmark}

\date{\today}

\begin{abstract}
Gravitational lensing by compact, small-scale intervening masses causes frequency-dependent distortions to gravitational-wave events.
The optimal signal-to-noise ratio (SNR) is often used as a proxy for the detectability of exotic signals in gravitational-wave searches.
In reality, the detectability of such signals in a matched-filtering search requires comprehensive consideration of match-filtered SNR, signal-consistency test value, and other factors.
In this work, we investigate for the first time the detectability of lensed gravitational waves from compact binary coalescences with a match-filtering search pipeline, GstLAL.
Contrary to expectations from the optimal SNR approximation approach, we show that the strength of a signal (i.e., higher optimal SNR) does not necessarily result in higher detectability.
We also demonstrate that lensed gravitational waves with wave optic effects can suffer significantly, from $~90\%$ (unlensed) to $<1\%$ (lensed) detection efficiency, due to downranking by the signal-consistency test values.
These findings stress the need to extend current template banks to effectively search for lensed gravitational waves and to reassess current constraints on compact dark matter scenarios.
\end{abstract}

\maketitle

\section{Introduction}\label{Section: Introduction}

Gravitational-wave detection typically relies on searching for signals using templates based on general relativity (GR)~\cite{Allen:2005fk}.
Commonly, templates used in the modeled search consider a simple quasicircular quadrupole mode waveform~\cite{Usman:2015kfa,LIGOScientific:2018mvr,Nitz:2018imz,Venumadhav:2019tad,Venumadhav:2019lyq,LIGOScientific:2020ibl,Nitz:2020oeq,LIGOScientific:2021usb,KAGRA:2021vkt,Andres:2021vew,Nitz:2021uxj,Nitz:2021zwj,Olsen:2022pin,Mehta:2023zlk,Kumar:2024bfe}.
While spin-orbit precessing effect, higher-order mode, and tidal deformability have been considered recently~\cite{Wadekar:2023gea,Schmidt:2024hac,Schmidt:2024jbp,Schmidt:2024kxy,Wadekar:2024zdq,Chia:2023tle},
there can be other physical effects such as orbital eccentricity, environmental effects and deviations from GR.
Within the GR framework, gravitational lensing emerges as the dominant mechanism for modifying gravitational waves during their propagation through space~\cite{Deguchi:1986zz,Nakamura:1997sw,Takahashi:2003ix}.
A distorted gravitational-wave signal, either by lensing or other means, can be problematic to detect with incomplete templates in modeled searches, potentially leading to missed detections or mischaracterizations of the source properties.
Gravitational lensing affects specific sources based on their alignment with an intervening mass, whereas non-GR effects generically alter the waveform of all gravitational wave sources regardless of their location or propagation path.
To determine whether the signals are detectable, one must know the probability that a random noise fluctuation produces the same effects in modeled searches, i.e., the false-alarm rate, which typical optimal SNR analyses cannot capture.

Gravitational waves can be deflected and distorted by intervening massive objects.
Consider the primary gravitational-wave(GW) sources for the ground-based detector, characterized by wavelength $\lambda_{\rm{GW}}\sim 10^6$~m, and a configuration in which the observer and the source are well aligned with the lens. 
Then, if the characteristic size of the lens $R_{\rm{Lens}}\gg \lambda_{\rm{GW}}$, such as galaxies ($R_{\rm{Lens}}\sim10^{20}$~m) or galaxy clusters ($R_{\rm{Lens}}\sim 10^{23}$~m), gravitational waves follow well-defined paths, and their magnification is achromatic~\cite{Wang:1996as, Dai:2017huk, Ezquiaga:2020gdt}.
This leads to distinct and well-separated (in time) copies of the original chirp that differ by a constant phase shift, relative magnification, and a relative time delay. 
This is the regime known as strong lensing, where multiple images are produced.
When $R_{\rm{Lens}} \sim \lambda_{\rm{GW}}$, lenses such as intermediate-mass black holes and dark matter clumps within a galaxy can cause a frequency-dependent modulation of the waveform.
This modulation, involving both interference and diffraction effects, can leave a lensing-induced distortion on a single signal~\cite{Deguchi:1986zz,Nakamura:1997sw,Takahashi:2003ix,Leung:2023lmq,Liu:2024xxn}. 
These distortions hold invaluable information about the possible population of small compact lenses~\cite{Jung:2017flg,Lai:2018rto,Christian:2018vsi,Dai:2018enj,Diego:2019lcd,Diego:2019rzc,Liao:2020hnx,Mishra:2021xzz,Liu:2023ikc,Chakraborty:2024mbr}, the caustic structure of different lenses~\cite{Lo:2024wqm}, the halo mass function~\cite{Fairbairn:2022xln,Tambalo:2022wlm,Caliskan:2022hbu,Savastano:2023spl}, and the cosmic expansion rate~\cite{Cremonese:2021puh,Chen:2024xal}. 
Understanding them is also key to avoid biases in parameter estimation~\cite{Cao:2014oaa,Liu:2024xxn} and false claims of violations of GR~\cite{Ezquiaga:2020spg,Gupta:2024gun,Liu:2024xxn}.

Since the LIGO-Virgo-KAGRA (LVK) Collaboration's \cite{LIGOScientific:2014pky,VIRGO:2014yos} first successful detection of gravitational waves in 2015, efforts have been continuously made to search for lensing signatures within gravitational-wave data~\cite{Hannuksela:2019kle,LIGOScientific:2021izm,LIGOScientific:2023bwz,Janquart:2023mvf,Janquart:2024ztv}. 
No significant evidence has been found yet.
Other efforts pointed out that no lensing features have been detected~\cite{McIsaac:2019use, Li:2019osa,Liu:2020par}, although 
Dai \textit{et al.}~\cite{Dai:2020tpj} also highlighted an intriguing candidate that is, however, astrophysically unlikely. 
These analyses are performed after detecting gravitational waves by different search algorithms designed to search for (unlensed) gravitational waves from compact binary coalescences (CBCs)~\cite{Allen:2005fk,LIGOScientific:2019hgc}.
The search algorithms use a predefined template bank to detect all the signals from CBCs~\cite{LIGOScientific:2018mvr,LIGOScientific:2020ibl,KAGRA:2021vkt}. 
However, as we will show in this work, signals can be missed if the lensing distortion is large. 
This can lead to biases in follow-up lensing analyses.
The probability of detecting a lensed signal thus may depend on the source and lens parameters, making it crucial to examine.
As the current templates lack gravitational lensing physics, it is fundamental to understand its impact on detectability in matched-filtering searches.

In this work, we study for the first time the detectability of lensed signals with state-of-the-art gravitational-wave search pipelines. 
We perform injection campaigns\footnote{In gravitational-wave searches, to test the search efficiency of a pipeline toward a certain type of gravitational-wave signals, we inject a set of simulated signals in real noise and assess how many signals the pipeline can recover.
This is known as an injection campaign.} to systematically study how lensing influences the detectability of gravitational-wave signals using a time-domain matched-filtering search pipeline GstLAL~\cite{2017PhRvD..95d2001M,Sachdev:2019vvd,2021SoftX..1400680C,Tsukada:2023edh}. 
GstLAL is one of the main pipelines the LVK Collaboration uses to search for gravitational-wave signals.
We examine the accuracy of the recovered source parameters, matched-filter SNRs, the signal-consistency test values, and the false-alarm rates (FARs) of the injections.

This paper is structured as follows:  Section~\ref{Section: Search} introduces the fundamentals of matched-filtering.
Section~\ref{gstlal} describes the essential parts of matched filtering in GstLAL.
Section~\ref{Section: Lensing} introduces gravitational lensing of gravitational waves using the point mass lens model and illustrates the expected effects of lensing on the detectability.
Section~\ref{Section: Methods} describes the injection sets and the setup of the matched filtering searches in our study.
Section~\ref{Section: Results} shows the results of the role of lensing and background noises in the detectability of lensed gravitational waves.
Section~\ref{Section: Conclusion} concludes the results and proposes new strategies for detecting lensed signatures in gravitational waves and underscores the implication of our work.

\section{Matched-Filtering Search for Gravitational waves}\label{Section: Search}

To access the detectability of gravitational-wave signals, one needs to predict their SNR in matched-filtering searches. Here we introduce the fundamentals of matched-filtering search and discuss the current efforts of detectability studies.

\subsection{Overview of matched-filtering search}
The expected gravitational-wave signals from CBC are well characterized by combinations of the binary component parameters.
Extrinsic parameters, such as source distances and inclinations, are not included in the template of matched-filtering searches. 
In the case of the matched-filtering search conducted in this study, which considers the intrinsic parameters, the parameter space is defined by the binary's component masses in the detector frame and spins. 
The component mass is denoted as $m_{i}$, where $i=\{1,2\}$, and by convention, $m_{1} \geq m_{2}$. 
The dimensionless spin is given as $\vec{\chi}_{i} = c \vec{S}_{i}/Gm^{2}_{i}$, where $c$ is the speed of light, $G$ is the gravitational constant, and $\vec{S}_{i}$ is the spin angular momentum.

The chirp mass determines the phase evolution during the inspiral phase of a binary system to the leading order~\cite{Peters:1964zz,Blanchet:1995ez},
\begin{align}
	\mathcal{M}_{c} = \frac{(m_{1}m_{2})^{3/5}}{(m_{1}+m_{2})^{1/5}}.
\label{chirp_mass}
\end{align}
The dimensionless spin $\vec{\chi}_{i}$ can combine to form an effective inspiral spin $\chi_{\rm{eff}}$~\cite{Ajith:2009bn,Santamaria:2010yb}, which is the leading contribution for spin in the post-Newtonian expansion and is defined as

\begin{align}
	\chi_{\rm{eff}} = \frac{(m_{1}\vec{\chi}_{1}+m_{2}\vec{\chi}_{2})\cdot \hat{L}_\mathrm{N} }{m_{1}+m_{2}},
\label{effective spin}
\end{align}
where $\hat{L}_\mathrm{N}$ is the unit vector in the direction of the Newtonian orbital angular momentum.
The spin parameters are assumed to be parallel to $\hat{L}_\mathrm{N}$ in our study.

Matched filtering is a powerful technique that is used to search for gravitational waves from CBCs and it is the optimal filter for detecting waveforms in stationary Gaussian noise~\cite{Allen:2005fk}. 
For a given search parameter space (e.g. in the component masses $m_1$-$m_2$ space), a set of gravitational waveforms $h_{T,\; j}$ (also known as ``templates") is precomputed and collectively stored in a ``template bank". The parameters of the templates are $\vec{\theta} = \{m1, m2, \chi_1, \chi_2\}$ and in the detector frame~\cite{Krolak:1987ofj}.

To identify a real gravitational-wave signal $h$, we compute the correlation between the template waveform $h_{T,\; j}$ and the data $d$ (which are composed of the background noise and potentially a signal) to construct the SNR $\rho$.
In the time domain, following~\cite{Allen:2005fk}, this is 
\begin{align}
    z_j(t) = x_j(t) + iy_j(t) &= 4\int_0^\infty \frac{\tilde{d}(f)}{\sqrt{S_n(f)}}\frac{\tilde{h}^*_{T,\; j}(f)}{\sqrt{S_n(f)}} e^{2 i \pi ft}df,
    \label{eq:SNR}
\end{align}
where $x_j(t)$ and $y_j(t)$ are the real and imaginary parts, respectively, of the complex SNR $z_j(t)$ using the $j$th template $h_{T,\;j}$.
The Fourier transforms of the time-domain signals $d(t)$ and $h_{T,\; j}$ are denoted as $\tilde{d}(f)$ and $\tilde{h}_{T,\; j}(f)$, and $*$ indicates the complex conjugate.
The square root of the noise power spectral density, $\sqrt{S_n(f)}$, is also called the noise amplitude spectral density.
The amplitude spectral density is used to ``whiten" both the data and the template waveform.
Finally, the factor of $4$ in Eq.~\ref{eq:SNR} is a normalization constant, such that 

\begin{align}
	1 = 4 \int^{\infty}_{0}\frac{|\tilde{h}_{T,\; j}(f)|}{S_n(f)}df.
\label{search1-1}
\end{align}
The SNR $\rho(t)$ is defined as the modulus of the complex SNR, i.e., $\rho(t) = |z_j(t)|$.

To quantify the loss in SNR due to an inexact model or to determine the fraction of missed signals in a matched-filter search, we compute the match, which is defined as
\begin{equation}\label{match_def}
    \mathcal{M}(h_{T,\; j}, h) = \underset{\phi_0, t_0}{\max} \frac{\langle h_{T,\; j}(\phi_0, \tau_0),h\rangle}{\sqrt{\langle h_{T,\; j}, h_{T,\; j}\rangle \langle h, h\rangle}},
\end{equation}

where $\langle\cdot,\cdot\rangle$, the noise-weighted inner product, is defined as
\begin{equation}
    \langle a,b\rangle = 4\mathrm{Re}\int_0^{\infty}\frac{\tilde{a}^*(f)\tilde{b}(f)}{S_n(f)}df.
\end{equation}

\begin{figure*}[t!]
\includegraphics[width=\textwidth]{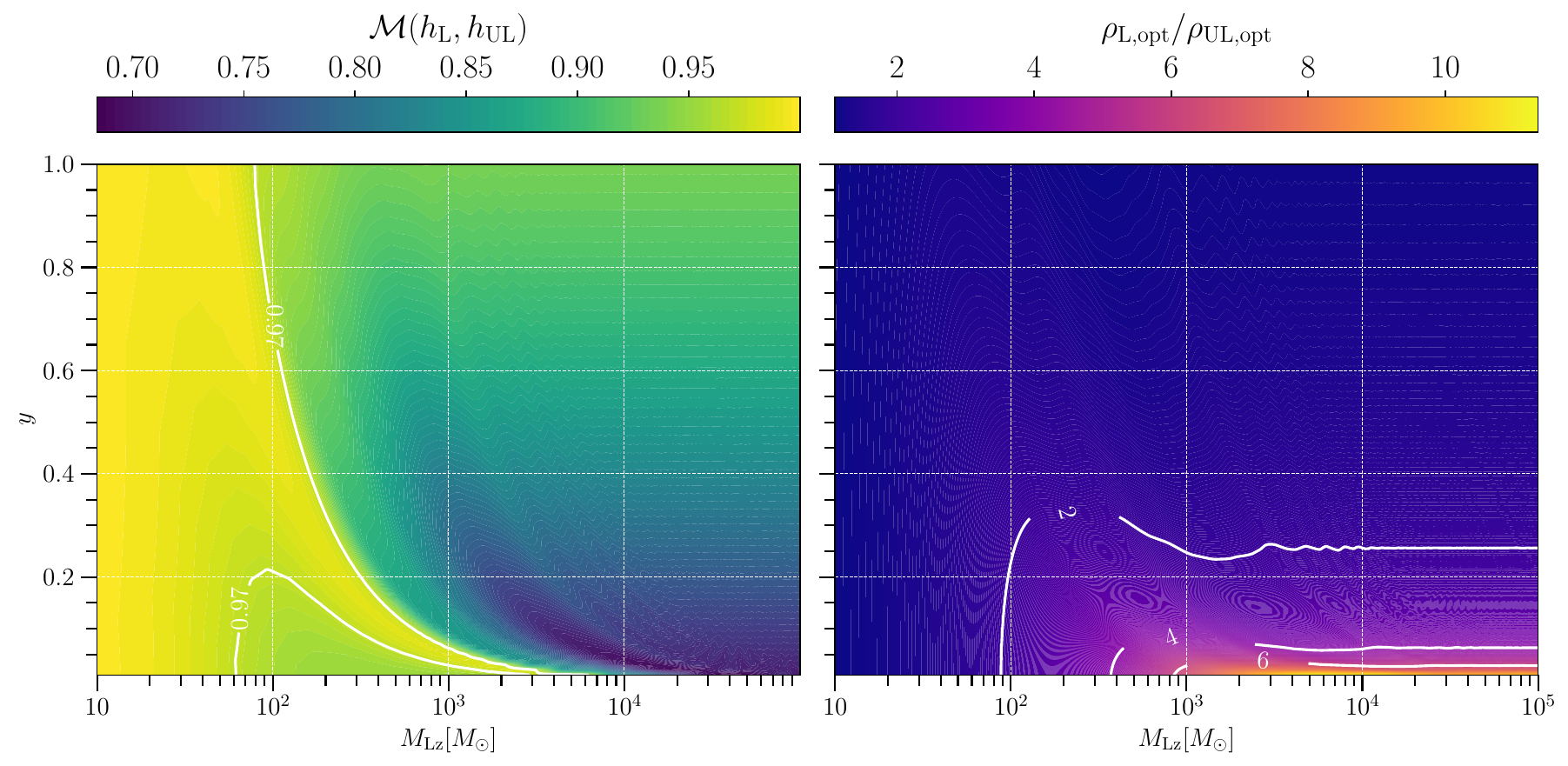}
\caption{\label{match_rho} Left: match between lensed signals and unlensed templates $\mathcal{M}(h_{\rm{L}}, h_{\rm{UL}})$ as a function of the redshifted lens mass $M_{\rm Lz}$ and impact parameter $y$. Right: ratio of the optimal signal-to-noise ratio of lensed $\rho_{\rm{L,opt}}$) and unlensed ($\rho_{\rm{UL,opt}}$) signals for the same $(M_{\rm Lz},y)$ parameter space). Here, we use a spinless binary black hole signal at 125~Mpc with $m_1 = m_2 = 30M_{\odot}$. $\rho_{\rm{L,opt}}/\rho_{\rm{UL,opt}}>1$ holds true for point mass lens model.} 
\end{figure*}

\subsection{Current efforts in detectability study}\label{effort}
In addition, the phase and time shifts, $\phi_0$ and $\tau_0$, maximize the match between the template waveform $h_{T,\; j}$ and the signal $h$. 
High $\mathcal{M}$ values indicate that $h_{T,\; j}$ and $h$ are well matched, where a maximum value of $\mathcal{M} = 1$ indicates that the signal is exactly described by the template waveform.
In the case where the template waveform exactly matches the signal (i.e. $h = h_{T,\; j}$) and the SNR is optimized with respect to noise realizations, the SNR is called the ``optimal SNR'' $\rho_{\rm opt}$ , and the ``observed SNR'' $\rho_{\rm obs}$ otherwise. 

Note that $\mathcal{M}$ is normalized by the waveform amplitude in Eq.~\ref{match_def}, whereas the SNR, as defined in Eq.~\ref{eq:SNR}, depends on both the waveform amplitude and waveform distortion.
Consequently, $\mathcal{M}$ is sensitive to additional physics that alter the waveform morphology but is insensitive to amplitude changes. 
On the other hand, the SNR is sensitive to both waveform distortion and changes in amplitude.

\begin{table*}[t!]
\begin{tabular}{cc|cccc|c}
\hline
\hline
Set & Type & Injection campaign & $y$ & $M_{\rm{Lz}}$ [$M_{\odot}$] & Source distance [Mpc] & Found injections (out of 19432)\\
\hline

1 & Unlensed & 1,2,3 & ... & ... & 125 & 17020 \\

2 & Lensed & 1 & 0.01 & $10$ & 125 & 17068 \\

3 & Lensed & 1 & 0.01 & $10^{3}$ & 125 & 14251 \\

4 & Lensed & 1,2 & 0.01 &  $10^{5}$ & 125 & 1 \\

5 & Unlensed & 2,3 & ... & ... & 625 & 12878\\

6 & Unlensed & 2,3 & ... & ... & 1250 & 6217\\

7 & Lensed & 2 & 0.01 & $10^{5}$ & 625 & 8\\

8 & Lensed & 2 & 0.01 & $10^{5}$ & 1250 & 13\\

9 & Lensed & 3 & $[0.01,1]$& $[10, 10^{5}]$ & 125 & 14303 \\

10 & Lensed & 3 & $[0.01,1]$ & $[10, 10^{5}]$ & 625 & 12536\\

11 & Lensed & 3 & $[0.01,1]$ & $[10, 10^{5}]$ & 1250 & 8254\\
\hline
\hline
\end{tabular}
\caption{\label{Table: injection sets} Set of 11 injections used in our three simulation campaigns. 
Injection campaign 1 studies the impact of lensing on detectability. 
Injection campaign 2 studies the role of signal strength. 
Injection campaign 3 gives a general picture of the detectability of the lensed gravitational waves. The found injection is defined as the FAR of the injection below the FAR threshold $3.85\times10^{-7}$ Hz (see Sec.~\ref{gstlal_lilkelihood} for FAR definition).
All injections used the $30-30M_{\odot}$ spinless binary black hole waveform (see Sec.~\ref{injections} for details on other source parameters).}
\end{table*}

Various methods, each with its own strengths and limitations, have been used to assess gravitational-wave signal detectability.
One widely used approach, employed in studies by~\cite{Tiwari:2017ndi,Veske:2021qis,Gerosa:2020pgy}, calculates the optimal SNR from a network of detectors, assuming that the same source parameters are recovered in matched-filtering searches.
This method has been applied to evaluate the detectability of gravitational-wave signals with wave optics effects, as seen in works by~\cite{Bondarescu:2022srx,Mishra:2023ddt}.
While this assumption is not strictly valid due to noise fluctuations, it is a useful proxy for assessing detectability, particularly in high-SNR scenarios.
Reference ~\cite{Essick:2023toz} improved this method by developing semianalytic model to estimate the sensitivity of gravitational-wave searches by analytically deriving the distribution of matched-filter SNR in Gaussian noise, incorporating effects like template bank correlations and network SNR maximization. 
This approach enables rapid predictions of detection probabilities without requiring large-scale injection campaigns, and offers a computationally efficient solution for the study of detection rates of CBC sources that require analyzing millions of signals.

In matched-filtering search pipelines, computation of SNR is followed by methods to mitigate the fluence of non-Gaussian noise flucuation and compute the ranking statistics.
For example, a signal consistency test has been employed to ensure the observed signals are consistent with the template bank.
If the gravitational-wave signal contains addition physics that deviate from the template, such as eccentricity, higher mode, non-GR effects, gravitational lensing, etc, it is regarded as inconsistent with the template and results in a lower detection efficiency.
Detailed discussion of the signal-consistency test with example is presented in Sec.~\ref{gstlal}.

Since matched-filtering search pipeline is critical in GW detections, an inacccurate modelling of the detection lead to selection biases of population constraints. 
For examples, ~\cite{Magee:2023muf} attempted to consider the effects of signal consistency test to investigate detectability of beyond-GR effects and
the selection biases induced and how they influence the population constraints.
For the most comprehensive analysis of detectability, complete injection campaigns are necessary,
which helps to investigate the impact of eccentricity~\cite{Ramos-Buades:2020eju} and higher-order modes~\cite{Capano:2013raa} on matched-filtering searches.
These campaigns simulate the detection process using search pipelines and simulated gravitational-wave signals.
This approach provides detailed information about the injected signals, including matched-filter SNR, signal-consistency test values, and final detection statistics.
While computationally intensive, injection campaigns offer the most accurate and complete assessment of gravitational-wave signal detectability across various scenarios and detector configurations~\cite{2017PhRvD..95d2001M}.

\section{The GstLAL search pipeline}\label{gstlal}
In this paper, we will focus on the matched-filtering search employed in the search pipeline, GstLAL. We will focus on the autocorrelation-based signal-consistency test detection statistic, and the ranking statistic assignment of GstLAL, which will aid our discussion in later sections of this paper.

\subsection{Autocorrelation-based signal-consistency test}
When $\rho(t)$ exceeds a predetermined threshold with $\rho=4$, the search pipeline registers the template parameters as a potential candidate, referred to as a trigger~\cite{2017PhRvD..95d2001M}. 
Non-Gaussianity and ``glitches" in the data can also produce high-SNR triggers that mimic gravitational waves. 
GstLAL therefore assigns for each trigger a signal-consistency test value $\xi^{2}$, which is based on the autocorrelation function. 
$\xi^2$ compares the difference between the expected autocorrelation time series of a template and the actual matched-filtered SNR time series. 
Mathematically, we have, for a given template $h_{T,\; j}$, 
\begin{align}
	\xi_{j}^{2} = \frac{\int_{-\delta t}^{\delta t} \left|z_{j}(t)-z_{j}(0)R_{j}(t)\right|^{2} dt}{\int_{-\delta t}^{\delta t} 2 - 2\left|R_{j}(t)\right|^{2}dt},
\label{search2}
\end{align}
where
\begin{align}
	R_j(t) = \int \frac{|\tilde{h}_{T, \; 2j}(f)|^2 + |\tilde{h}_{T, \; 2j+1}(f)|^2}{S_n(f)} e^{2\pi i f t} df,
\label{search2-1}
\end{align}
is the expected SNR time series. Here $t=0$ is the time of the global maximum in the actual SNR time series. The tunable parameter $\delta t$ defines the size of the window around the peak time over which to perform the integration~\cite{2017PhRvD..95d2001M}. 
$\xi_j^2$ is normalized so that a perfect alignment gives $1$. Deviation from the expected SNR time series will result in larger $\xi^{2}$.

Consider the extreme case where the trigger from a glitch is a scaled $\delta$ function $z_j(0)\delta(t)$ at time $t=0$ with amplitude $z_j(0)$. 
The actual SNR time series will then be identically $0$ everywhere over the integration range except for $t=0$, which differs from the expected SNR time series [i.e. $z_j(0) R_j(t)$] significantly. 
Essentially, the numerator of the $\xi^2$ will become large. 
In fact, the larger the SNR of the noise trigger [i.e. the larger the value of $z_j(0)$], the higher the value $\xi^2$ will be.
Also, a signal that differs from the template waveforms can potentially impact the $\xi^{2}$ and is not trivial~\cite{Bondarescu:2022srx}.

\begin{figure*}[t!]
\includegraphics[width=\textwidth]{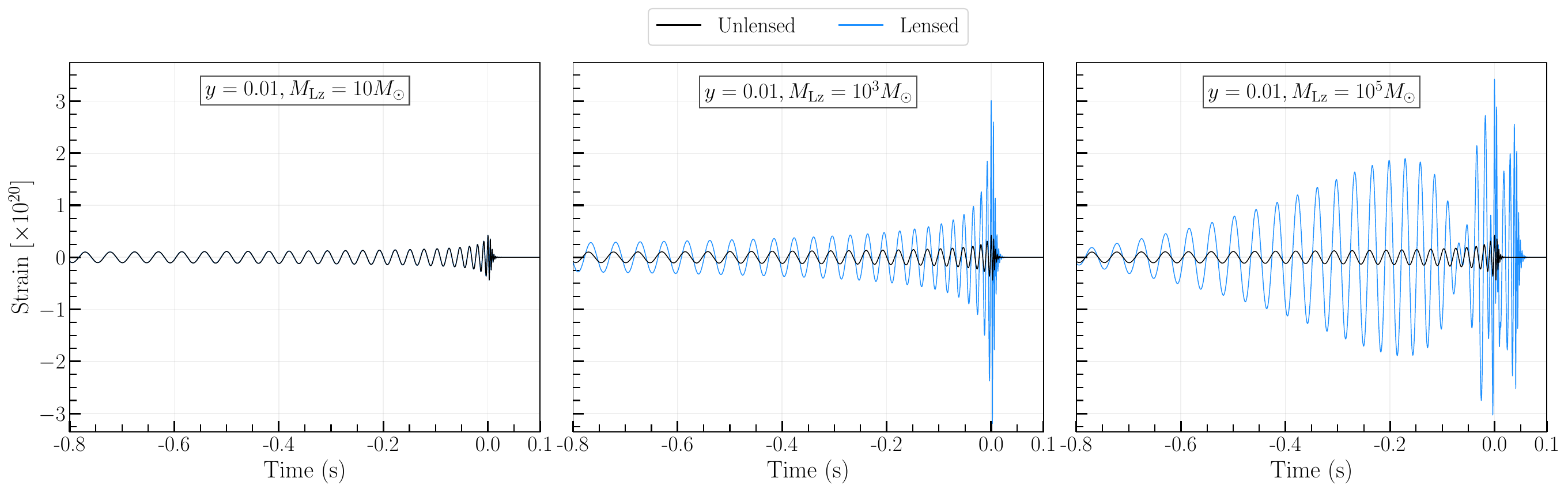}
\caption{\label{PML_WF} Time-domain gravitational-wave waveforms of a spinless binary black hole with component masses $m_1=m_2=30 M_{\odot}$ in unlensed (black) and lensed (blue) scenarios that represent increasing amplification effects. The lensed waveform has the same impact parameter $y=0.01$. Left: redshifted lens mass $M_{\rm{Lz}} = 10M_{\odot}$ with match $\mathcal{M} = 0.99$. Middle: $M_{\rm{Lz}} = 10^3 M_{\odot}$, $\mathcal{M} = 0.97$. Right: $M_{\rm{Lz}} = 10^5 M_{\odot}$, $\mathcal{M} = 0.7$. Lensing distortion can enlarge the amplitude.} 
\end{figure*}

\subsection{Likelihood-ratio ranking statistic}\label{gstlal_lilkelihood}
For each trigger, GstLAL assigns a likelihood-ratio ranking statistic based on the detection statistics (e.g. $\rho$, $\xi^2$)~\cite{2017PhRvD..95d2001M,Sachdev:2019vvd,2015arXiv150404632C,Fong:2018elx}. 
The likelihood ratio is the ratio of probabilities of obtaining the set of detection statistics for a trigger under the signal hypothesis to that under the noise hypothesis. 
Mathematically, it is defined as 
\begin{align}
	\mathcal{L} = \frac{P(\vec{D}_H,\vec{O},\boldsymbol{\vec\rho,\vec\xi^2},\left[\vec{\Delta t}, \vec{\Delta \phi}\right]|\vec\theta,\mathbf{signal})}{P(\vec{D}_H,\vec{O},\boldsymbol{\vec\rho,\vec\xi^2},\left[\vec{\Delta t}, \vec{\Delta \phi}\right]|\vec\theta,\mathbf{noise})}\cdot\frac{P(\vec\theta|\mathbf{signal})}{P(\vec\theta|\mathbf{noise})},
\label{likelihood_ratio_ranking}
\end{align}
where $\vec{D}_H$ are the horizon distances of the detectors,  $\vec{O}$ is the set of detectors that registered the trigger, $\vec\rho$ are the recorded detector SNRs for the trigger at each participating detector, $\vec\xi^2$ are the autocorrelation-based consistency test values computed for each of the participating detectors, and $\left[\vec{\Delta t}, \vec{\Delta \phi} \right]$ are the sets of differences in times $\Delta t$ and coalescence phases $\Delta \phi$ measured by the participating detectors.

In previous works~\cite{Tiwari:2017ndi,Veske:2021qis,Gerosa:2020pgy}, it has been assumed that optimal SNRs can be used as a proxy to determine the detectability of gravitational-wave signals in matched-filtering-based detection pipelines. However, we remind readers that such an assumption is incorrect. It has been shown that the likelihood ratio provides the most powerful detection statistic at a given false-alarm probability ~\cite{Neyman:1933wgr,Cannon:2015gha}. Matched-filtered SNR will be a good proxy if it scales approximately linearly with the likelihood ratio, which can only be true if the noise in gravitational-wave data is stationary and Gaussian. Therefore, a thorough injection study is required to explore the detectability of gravitational-wave signals instead of using matched-filtered SNRs as a proxy. 

GstLAL then computes the complementary cumulative distribution defined by
\begin{align}
    \mathcal{C}(\ln \mathcal{L}^* | \text{noise}) = \int_{\ln \mathcal{L}^*}^\infty P\left(\ln \mathcal{L} | \text{noise}\right) d\ln\mathcal{L},
\end{align}

which allows us to compute the false-alarm probability (FAP, or the p value) as the probability that $M$ independent coincident noiselike events ($\{N_1, N_2, ... , N_M\}$ contain at least one event with a log likelihood ratio $\mathcal{L}$ larger or equal to some threshold $\ln\mathcal{L}^*$:

\begin{align}
    P(\ln\mathcal{L}\geq \ln\mathcal{L}^* | N_1, ... ,N_M) &= 1 - \left(e^{-\mathcal{C}(\ln \mathcal{L}^* | \text{noise})}\right)^M\\
    &= 1 - e^{-M\mathcal{C}(\ln \mathcal{L}^* | \text{noise})}
\end{align}

where $e^{-\mathcal{C}(\ln \mathcal{L}^* | \text{noise})}$ denotes the probability that a Poisson process with mean rate $\mathcal{C}(\ln \mathcal{L}^* | \text{noise})$ can yield an event with log likelihood ratio $\mathcal{L} < \ln \mathcal{L}^*$.
The FAP denotes the probability that noise can produce a trigger with likelihood ratio $\mathcal{L}$ larger or equal to the likelihood ratio $\mathcal{L}^*$ of the trigger under consideration. One would expect triggers from real gravitational-wave signals to have higher likelihood ratios. 
Hence, the lower the FAP, the more likely the candidate is a gravitational wave.
A similar quantity known as the false-alarm rate (FAR), is also evaluated for each candidate. FAR is defined as
\begin{align}\label{FAR}
    \text{FAR} = \frac{\mathcal{C}(\ln\mathcal{L}^*|\text{noise})}{T}
\end{align}
where $N$ is the total number of triggers and $T$ is the total observing time.
Essentially, FAR reflects the rate of noise producing a trigger with likelihood ratio $\mathcal{L}$ larger or equal to that of the trigger under consideration $\mathcal{L}^*$.
The lower the FAR, the more likely the candidate is a real gravitational-wave signal~\cite{2017PhRvD..95d2001M,Sachdev:2019vvd,2015arXiv150404632C,Fong:2018elx}.
The typical FAR threshold to claim a confident GW detection in GstLAL is $3.85\times10^{-7}$ Hz ($1$ in $30$ days).

\section{Brief introduction to Gravitational Lensing}\label{Section: Lensing}

In the following sections, we introduce the theory of gravitational lensing and show how lensing imprints frequency-dependent modulation on the waveforms using the point mass lens model.
This modulation includes interference and diffraction, resulting in waveform distortion and amplitude enhancement.
We briefly discussed the effects of lensing distortion on detectability in mathced filtering, using optimal SNR as a proxy.

\begin{figure*}[t!]
\includegraphics[width=\textwidth]{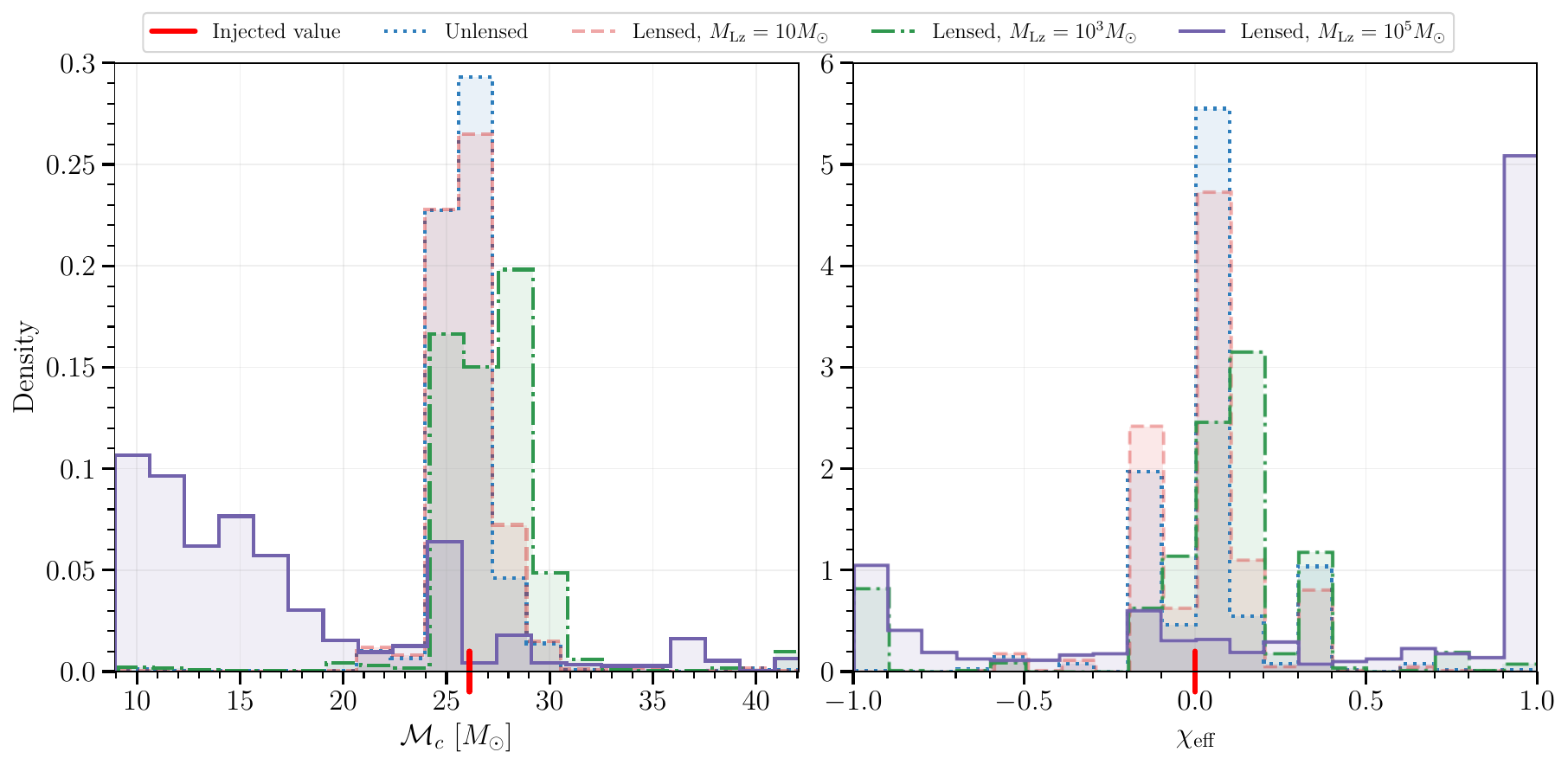}
\caption{\label{Fig: set1_recovered_template} The role of lensing in the accuracy of the recovered source parameters of lensed gravitational waves. The panels show the histogram of the recovered source parameter with different lens parameters in our first injection campaign (see Table~\ref{Table: injection sets} and Fig.~\ref{PML_WF}). Left: chirp mass $\mathcal{M}_c$. Right: effective spin $\chi_{\rm{eff}}$. The red solid line on the x-axis represents the injected value of the $30-30$~$M_{\odot}$ spinless binary black hole signals at 125~Mpc. The unlensed injections are blue. Lensed injections with the impact parameter $y=0.01$ and redshifted lens mass $M_{\rm{Lz}} = 10M_{\odot}$ with match $\mathcal{M}=0.99$ (red), $10^{3}M_{\odot}$, $\mathcal{M}=0.97$ (green), $10^{5}M_{\odot}$, $\mathcal{M}=0.7$ (purple). Lensing distortion can lead to a substantial reduction in the accuracy of the source parameters.}
\end{figure*} 

\subsection{Theory of gravitational lensing of gravitational waves}

Gravitational lensing theory describes how waves propagate in a curved background. 
When considering the gravitational-wave propagation in the weak-field limit, where the Newtonian gravitational potential $U\ll 1$ and employing the thin-lens approximation, gravitational lensing is well described by the Fresnel-Kirchhoff diffraction integral over the whole lens plane in the frequency domain~\cite{Deguchi:1986zz,Nakamura:1997sw,Takahashi:2003ix},

\begin{equation}
F(f, \pmb{y}) = \frac{f}{i}\int d^2\pmb{x} ~ e^{-i2\pi f \tau_{\rm d}(\pmb{x}, \pmb{y})} ,
\label{fres}
\end{equation}

\noindent where $f$ is the frequency of the gravitational waves.
$\pmb{x}\equiv(x_1, x_2)$ and $\pmb{y}\equiv(y_1, y_2)$ represent the image plane and source plane coordinates, respectively.
The time delay function $\tau_{\rm d}(\pmb{x}, \pmb{y})$ is given as

\begin{equation}
    \tau_{\rm d}(\pmb{x}, \pmb{y}) = \frac{(1+z_{\rm L})\xi_0^2}{c}\frac{D_{\rm s}}{D_{\rm d} D_{\rm ds}} 
    \left[\frac{1}{2}(\pmb{x} -\pmb{y})^2 - \psi\left(\pmb{x}\right)
    + \phi_{\rm m}\left(y\right)\right]\,,
\end{equation}

where $z_{\rm L}$ is the lens redshift.
$D_{\rm s}$, $D_{\rm d}$,  and $D_{\rm ds}$ are the angular diameter distances between the observer and the source, the observer and the lens,  and the lens and the source respectively. 
$\phi_{m}$ is the minimum value of the Fermat potential at a given $y$.
$\xi_0$ is an arbitrary reference length unit on the lens plane used to make the coordinates dimensionless.

Lensing of gravitational waves is governed by frequency-dependant amplification factors $F(f)$~~\cite{Takahashi:2003ix}.
In the frequency domain, lensed and not lensed gravitational-wave signals ($h_{\rm{L}}$ and $h_{\rm{UL}}$) are related by
\begin{equation}
h_{\rm{L}}(f)\!=\!F(f,\boldsymbol{y}) h_{\rm{UL}}(f)~, \label{eq:lensedgravitational wave}
\end{equation}

The characteristic wavelength of the CBC sources sensitive the ground-based detector is $\lambda_{\rm{GW}}\sim10^{6}$~m.
Lenses like stars, stellar remnants, star clusters and compact dark halos in a galaxy, have $R_{\rm{Lens}}\sim \lambda_{\rm{GW}}$.
This corresponds to the redshifted lens mass $M_{\rm Lz} \equiv M_{\rm L}(1+z_{\rm{L}})=[10,10^{5}] M_{\odot}$, where $M_{\rm{L}}$ is the lens mass in the lens frame.
In this regime, calculating the amplification factors with Eq.(~\ref{fres}) becomes necessary.
We refer to it as a wave optics regime in the rest of our text.

In this paper, we adopt isolated point mass as our lens model, which has two lens parameters: $M_{\rm Lz}$  and dimensionless source position $y$~\footnote{We use the conventional notation $y=\eta / \theta_{\rm{E}}$, where $\eta$ is the displacement between a source and line of sight and $\theta_{\rm{E}}$ is the Einstein radius of a lens.}.

The amplification factor for a point mass lens is given by
\begin{align}
\label{pmaf}
    F(w,y)&=\bigg{[}\exp\bigg{[}\dfrac{\pi w}{4}+i \dfrac{w}{2} \bigg{\{}\text{ln} \bigg{(}\dfrac{w}{2}\bigg{)}-2\phi_m(y)\bigg{\}}\bigg{]} \nonumber \\
&\times \Gamma \bigg{(}1- \dfrac{i}{2}w\bigg{)} {_1}F_1\bigg{(}\dfrac{i}{2}w, 1; \dfrac{i}{2}w y^2\bigg{)}\bigg{]}^*,
\end{align}
where $\omega \equiv 8\pi M_{\rm{Lz}} f$ is the dimensionless frequency, and ${_1}F_1$ is the hypergeometric function~~\cite{Deguchi:1986zz,Nakamura:1997sw,Takahashi:2003ix}.

We set the ranges of the two lens parameters as $(M_{\rm Lz},y) = ([10,10^{5}]M_{\odot},[0.01,1.0])$ to cover regions of the long-wavelength regime, wave-dominated regime, and geometric-optics regime.
For $M_{\rm Lz} < 10M_{\odot}$, the resulting lensed waveform is similar to the unlensed waveform.
For $M_{\rm Lz} > 10^{5}M_{\odot}$, the lensed waveform is in the geometric-optics regime, and there are distinct images only with an overall amplification and constant phase shift.
Additionally, we exclude cases where $y>1.0$ because lensing distortions are comparatively weaker.

\subsection{The impact of lensing on signal detectability in matched-filtering searches: Match and optimal SNR}

To estimate how lensing within the given parameter space affects the waveform morphology, we calculate the match between the lensed $\mathcal{M}(h_{\rm{L},}h_{\rm{UL}})$ and intrinsic unlensed waveform and the ratio of the optimal SNR in each lensed and unlensed scenario, $\rho_{\rm{L,opt}}/\rho_{\rm{UL,opt}}$, where $\rho_{\rm{L,opt}}$ is the optimal SNR between the lensed waveforms and $\rho_{\rm{UL,opt}}$ is the optimal SNR between the unlensed waveforms, is considered.
A $30$-$30~M_{\odot}$ spinless waveform template is used in both computations. 
As discussed in Sec.~\ref{effort}, $\mathcal{M}(h_{\rm{L}}, h_{\rm{UL}})$ is normalized by both the lensed and unlensed waveform amplitudes and is hence insensitive to lensing magnification.
In contrast, $\rho_{\rm{L,opt}}$ accounts for both waveform distortion and amplitude changes. 
Hence, if the lensing magnification effect dominates over the waveform distortion effect, we have $\rho_{\rm{L,opt}}/\rho_{\rm{UL,opt}} > 1$.

As shown in the left panel of Fig.~\ref{match_rho}, the match in the region with high $M_{\rm Lz}$ and low $y$ values is remarkably low.
This occurs because the corresponding lensed signals exhibit distinctive beating patterns due to high magnifications and short time delays, as depicted in the right panel of Fig.~\ref{PML_WF}.
Thus, using unlensed waveforms as templates to analyze the lensed waveforms can lead to inaccurate results.

Furthermore, the optimal SNR assumes that the best-fit template is the same as the signal.
A higher optimal SNR value generally indicates greater signal detectability.
If a signal is lensed by a point mass in a wave optics regime, it is always magnified, resulting in a higher optimal SNR value.
The right panel in Fig.~\ref{match_rho} illustrates how optimal SNR values increase depending on the lens parameters.
As expected, pairs with high $M_{\rm Lz}$ and low $y$, corresponding to higher magnifications, lead to higher optimal SNRs for lensed signals.
The right panel of Fig.~\ref{PML_WF} explains the difference between the left and the right panel in Fig.~\ref{match_rho}, 
An isolated point mass can enhance the waveform amplitude by at least five times while $\mathcal{M}(h_{\rm{L},}h_{\rm{UL}})=0.7$.
It represents a scenrio where the lensing magnification overwhelms the waveform distortion, result in $\rho_{\rm{L,opt}}/\rho_{\rm{UL,opt}} \sim 10$.
This is qualitatively consistent with the matched-filter analysis in \cite{Mishra:2023ddt}.

To distinguish the lensing distortion from glitches or deviations from general relativity~\cite{Gupta:2024gun,Chandramouli:2024vhw,Liu:2024xxn}, one needs to perform parameter estimation and Bayesian model analysis, which are beyond the scope of this paper, interested reader please find a systematic study ~\cite{Gupta:2024gun,Chandramouli:2024vhw,Liu:2024xxn} to study the systematics between lensing and deviation from general relativity.
These analyses are performed based on a confirmed detection from the search pipeline and are therefore subject to bias if the detectability of the signal is not well understood.

In matched-filtering searches, the detectability of gravitational-wave signals is approximated by observed SNRs that can be estimated using optimal SNR.
In real detection scenarios, detector noise fluctuates over time, which can cause a template with different source parameters, $h_{\rm tem}$, to be the best-fit template for recovering the detected gravitational-wave waveform, $h_{\rm true}$. As a result, the observed SNR and optimal SNR values can differ from the values obtained in the absence of noise.
In addition to noise effects, lensing  (i.e., waveform distortion) can further affect the search pipeline's ability to recover the best-fit template, resulting in a nontrivial behavior in the observed and optimal SNR.
Therefore, one need to perform an injection campaign to understand how gravitational lensing influence the detectability in matched-filtering searches.

\begin{figure*}[t]
\includegraphics[width=\textwidth]{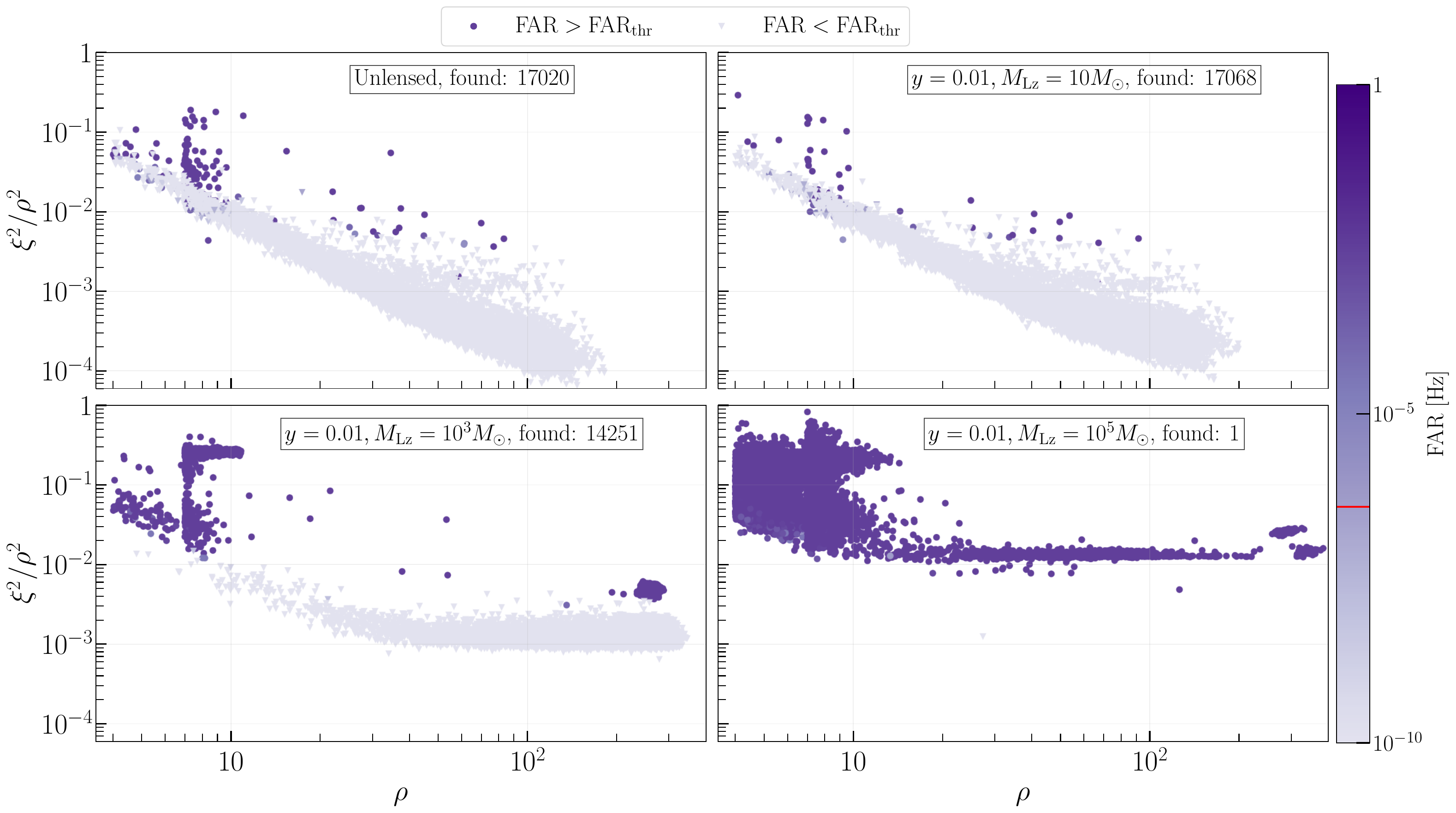}
\caption{\label{Fig: set1_chisq_snr_far} 
The role of lensing in the detectability of lensed gravitational waves. The panels show the distribution of the simulated events in our first injection campaign (see Table~\ref{Table: injection sets} and Fig.~\ref{PML_WF}) as a function of matched-filter SNR $\rho$ and the signal-consistency test $\xi^{2}/\rho^{2}$ [see Eq.(~\ref{search2})]. The color bar represents the FAR. An event is considered a significant gravitational-wave candidate when its FAR is below $3.85\times10^{-7}$ Hz or one every 30 days, indicated by the red tick. The circle and triangular markers represent larger and smaller injections than the FAR threshold for the detection, respectively. Upper left: unlensed injections. The remaining panels are lensed injection with impact parameter $y=0.01$. Upper right: redshifted lens mass $M_{Lz}=10M_{\odot}$ with match $\mathcal{M}=0.99$. Bottom left: $M_{Lz}=10^3M_{\odot}$, $\mathcal{M}=0.97$. Bottom right: $M_{Lz}=10^5M_{\odot}$, $\mathcal{M}=0.7$. Lensing distortion can introduce a significant drop in $\rho$ and a larger $\xi^{2}$ value in matched-filtering searches}
\end{figure*}

\section{Methodology}\label{Section: Methods}

We perform three injection campaigns using a time-domain matched-filtering search pipeline, GstLAL, to investigate how lensing impacts the detectability of gravitational-wave signals in searches.
The distributions of the recovered source parameters, matched-filter SNR, the signal-consistency test value, and the FAR inform the waveform deviation and noise fluctuations in matched-filtering searches, which explain changes in detectability  when varying the source parameters.
The details of the matched-filtering search setup and injection parameters are described in Table~\ref{Table: injection sets}.

\subsection{Matched-filtering searches}\label{method_search}

In matched-filtering searches, accurate likelihood-ratio calculations for the injections require collecting sufficient noise background to evaluate the denominator in Eq.~(\ref{likelihood_ratio_ranking}).
Moreover, templates with different source parameters can recover the same injections due to noise fluctuations.
These two conditions require a template bank with a parameter space sufficiently larger than the injections' source parameters. 
We use representative $30$-$30$ solar masses and a spinless binary black hole (BBH) as our source parameters, so the only key independent variables are the lensing parameters (strength of lensing) and source distances (strength of signals).
This allows us to construct a smaller template bank to search for gravitational waves efficiently in our study following~\cite{Privitera:2013xza}. The bank covers BBH masses with $m_{1}= [10M_{\odot}, 90M_{\odot}]$, $m_{2}= [10M_{\odot}, 40M_{\odot}]$ and includes 14606 templates, ensuring that the injection lives within the parameter space covered by this bank. 
Similar to the third Gravitational-Wave Transient Catalog GWTC-3, we use \\\textsc{seobnrv4\_rom} waveform model approximant~\cite{Bohe:2016gbl} for our templates with pure quadrupole mode and aligned spin~\cite{LIGOScientific:2021usb}.
The aligned component spins are distributed in the range $[-0.99, 0.99]$.
One potential caveat is the accuracy of the FAR computation when using a smaller template bank for the detectability study.
As demonstrated in the results of the first injection campaign (Sec.~\ref{sec:result_A}), the FAR computation in our detectability study is primarily affected by the accuracy of the recovered source parameters, which means that a larger template bank simply permits greater scope for inaccurate recovery of source parameters at a higher computational cost.
This justifies our use of a smaller template bank.

We analyzed LIGO Hanford and Livingston data from the third observing run (O3) of the LVK Collaboration.
Our data include real noises and spans from the first week of O3 with GPS time $1238166011$-$1238787954$~s. 
The data quality and glitch mitigation follows that of GWTC-3 ~\cite{LIGOScientific:2021usb}.
To assess the impact of lensing on matched-filter searches, we analyzed the distributions of recovered chirp mass, effective spin, matched-filter SNR, the autocorrelation-based signal-consistency test value $\xi^{2}$ in GstLAL, and the resultant FAR.
The detection threshold for our searches is set as the representative $\rm{FAR}_{\rm{thr}} = 1$ in $30$ days.

\begin{figure*}[t!]
\includegraphics[width=\textwidth]{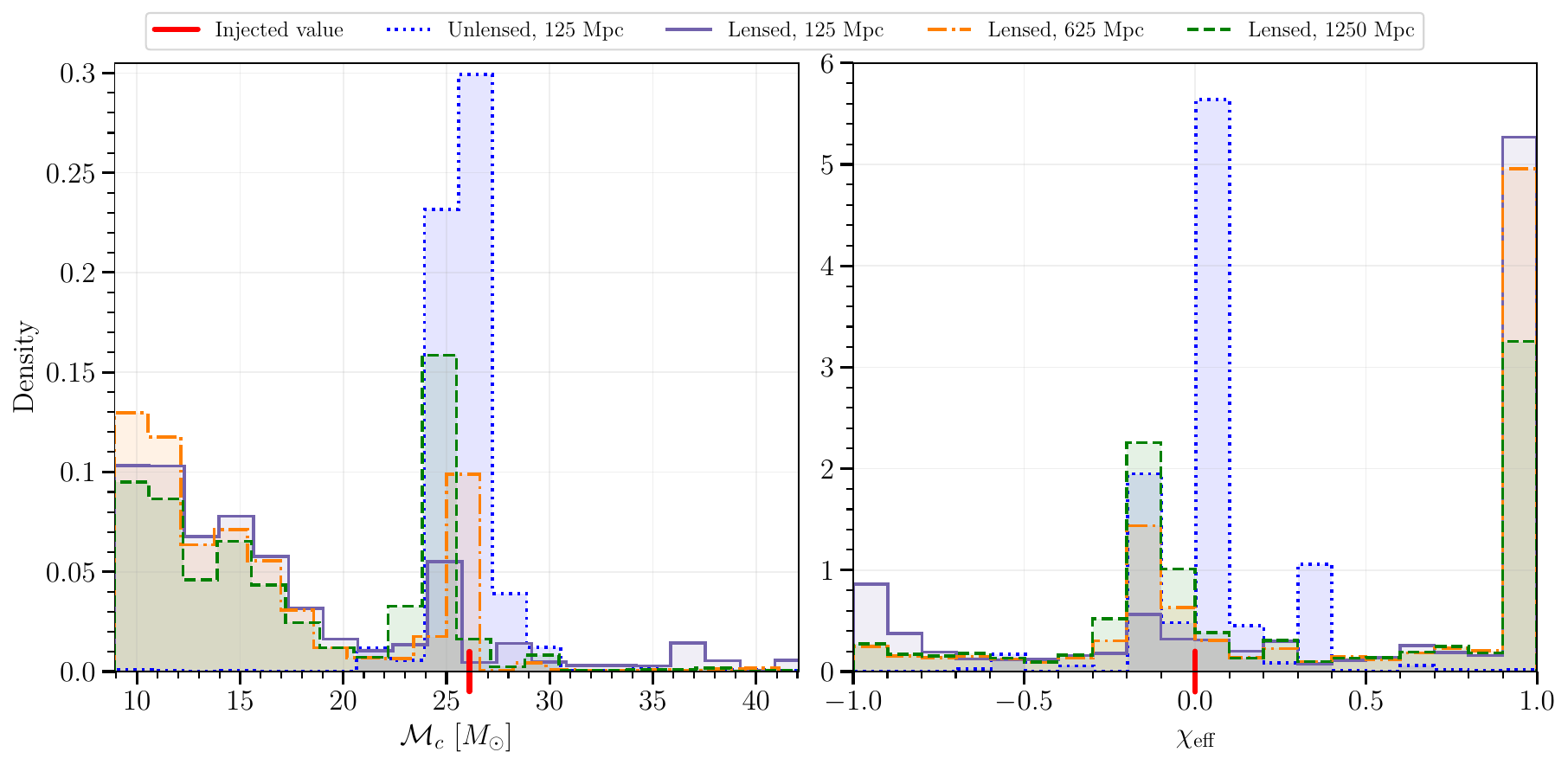}
\caption{\label{Fig: set2_recovered_template} The role of signal strength in the accuracy of the recovered source parameters of lensed gravitational waves. The panels show the histogram of the recovered source parameter at different source distances in our second injection campaign (see Table~\ref{Table: injection sets}). Left: chirp mass $\mathcal{M}_c$. Right: effective spin $\chi_{\rm{eff}}$. The red solid line on the x-axis represents the injected value of the $30-30$~$M_{\odot}$ spinless binary black hole signals. The unlensed injections are at 125 Mpc (blue). All lensed injections use impact parameter $y=0.01$ and redshifted lens mass $M_{\rm{Lz}}=10^{5}M_{\odot}$ at difference source distances: 125 (purple), 625 (orange), 1250 Mpc (green). The recovered source parameter is more accurate at larger source distances (lower SNR) when the lensing distortion is significant.}
\end{figure*}

\subsection{Injection campaigns}\label{injections}

We first conduct two injection campaigns to study how lensing and signal strength influence detectability.
These allow us to predict the general picture of how lensing can impact the detectability of gravitational waves in matched-filtering searches in the third injection campaign. 
We prepare $11$ injection sets, each consisting of $19432$ injections.
All injections are generated using the \\\textsc{seobnrv4}~\cite{Bohe:2016gbl} waveform approximant in \texttt{LALSimulation}~\cite{lalsuite}. 
We uniformly sample over the right ascension, declination, inclination, and polarization for all injections.
All injections are distributed uniformly across the data over time.
We summarize the lens parameters, the source distances, and the number of found injections in Table~\ref{Table: injection sets}. 
In the following sections, we will discuss how we systematically vary the lens parameters and source distances for different injection sets to compare the role of lensing and signal strength on the detectability.

\subsubsection{Injection campaign 1: The role of lensing}\label{inj1}
The first injection campaign explores the impact of lensing amplification on gravitational-wave signal detectability.
Four injection sets are examined: one unlensed set and three lensed sets with varying lensing parameters ($y=0.01$ and $M_{\rm{Lz}}=\{10, 10^{3}, 10^{5} \}M_{\odot}$).
These sets represent different lensing regimes: negligible amplification (long wavelength), significant amplification without beating patterns (intermediate), and wave dominated with notable beating patterns, as shown in Fig.~\ref{PML_WF}.
As lensing strength increases across sets, the match between lensed and unlensed waveforms decreases ($\mathcal{M}=\{0.99, 0.98, 0.7\}$).
As shown in Fig~\ref{match_rho}, conventional optimal SNR analysis suggests stronger gravitational lensing, which increases optimal SNR and should enhance signal detectability, which will be examined in this injection campaign.

\subsubsection{Injection campaign 2: The role of signal strength}\label{inj2}
The first injection campaign utilized a fixed unlensed SNR of $\rho \sim 100$, representing an unrealistically high-SNR regime with O3 detector sensitivity and minimal impact from background noise.
To better understand how signal strength influences lensing's impact on detectability, a second injection campaign was conducted.
This campaign used the same lens parameters as set 4, the one giving the largest lensing distortions ($y=0.01$ and $M_{\rm{Lz}}=10^{5} M_{\odot}$), and compared it to the unlensed case by varying source distances from $d_{L}=\{125, 625, 1250\}$~Mpc.
This resulted in injections with unlensed optimal SNRs of $\rho_{\rm{opt}} \sim \{100, 50, 10 \}$ at O3 sensitivity, corresponding to injection sets 1 and $4$-$8$ in Table~\ref{Table: injection sets}.
Additionally, this investigation hints at whether improved detector sensitivities, leading to higher detector SNRs, could enhance the detectability of lensed gravitational waves.

\begin{figure*}[t!]
\includegraphics[width=\textwidth]{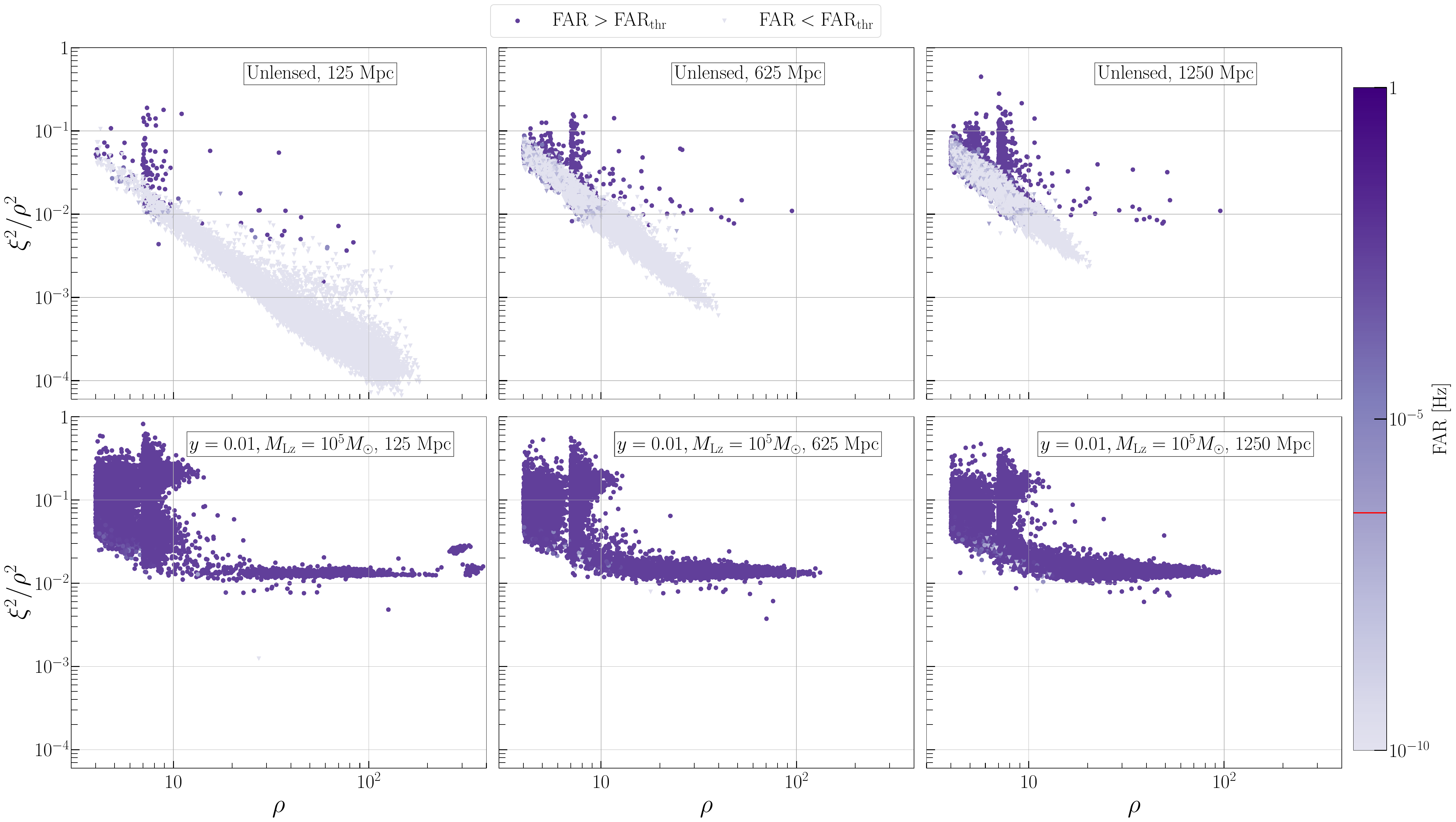}
\caption{\label{Fig: set2_chisq_snr_pass_lost} The role of signal strength in the detectability of lensed gravitational waves. The panels show the distribution of the simulated events in our second injection campaign (see Table~\ref{Table: injection sets}) as a function of matched-filter SNR $\rho$ and the signal-consistency test $\xi^{2}/\rho^{2}$ (see Eq.~\ref{search2}). The color bar represents the FAR. An event is considered a significant gravitational-wave candidate when its FAR is below $3.85\times10^{-7}$ Hz or one every 30 days, indicated by the red tick. The circle and triangular markers represent larger and smaller injections than the FAR threshold for the detection, respectively. Upper: unlensed injections. \textit{Bottom}: lensed injections with impact parameter $y=0.01$,and redshifted lens mass $M_{\rm Lz}=10^5M_{\odot}$. \textit{Left}: $125$~Mpc. Middle: $625$~Mpc. Right: $1250$~Mpc. Gravitational waves with strong lensing distortion are less detectable at lower source distances (higher SNR).}
\end{figure*}

\subsubsection{Injection campaign 3: detectability of lensed gravitational waves}\label{inj3}
The initial two injection campaigns elucidated the roles of lensing and signal strength, enabling us to predict lensed gravitational-wave detectability more comprehensively.
To present a general picture of how lens parameters influence matched-filtering searches, we sampled over $y\in[0.01,1]$ and $M_{\rm{Lz}}\in[10M_{\odot},10^{5}M_{\odot}]$ for each injection set. 
We further verified the optimal SNR analysis by comparing matched-filter SNRs of lensed and unlensed injections, using sets with $d_{L}={125, 625, 1250}$ Mpc (corresponding to sets 1, 5, 6, and $9$-$11$ in Table~\ref{Table: injection sets}).
These six injection sets also provide insights into the potential need for extending current template banks to include lensed gravitational waves by comparing their detectability in matched-filtering searches across different SNRs.

The first two injection campaigns (sets $1$-$8$) used fixed component masses, spins, lens parameters, and source distances for each set.
Distributing these injections across different times in the data allowed for systematic comparison of recovered $\mathcal{M}_c$ and $\chi_{\rm{eff}}$ under various noise realizations, providing specific insights into how lensing and signal strength impact matched-filter SNR.
The third campaign, sampling over lens parameters, sacrificed these specific details for a more general picture of lensing's impact on detectability, as each injection had different noise realizations and lens parameters.
Consequently, recovered $\mathcal{M}_{c}$ and $\chi_{\rm{eff}}$ are presented only for the first two campaigns, while matched-filter SNR, $\xi^{2}$, and FAR are reported for all campaigns.

\begin{figure*}[t!]
\includegraphics[width=\textwidth]{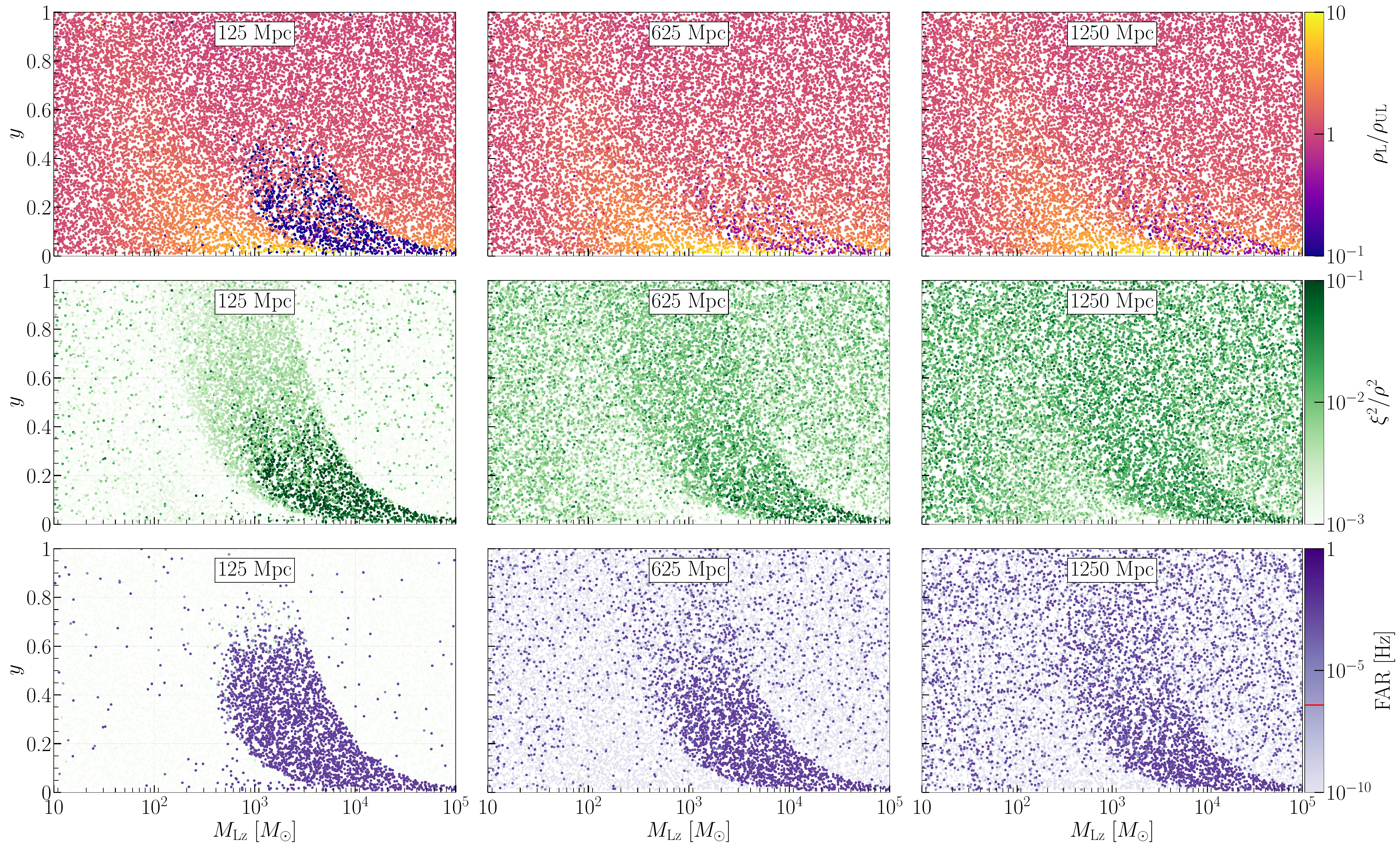}
\caption{\label{set3_all} Detectability of lensed gravitational waves.
The panels show the distribution of the simulated events in our third injection campaign (see Table~\ref{Table: injection sets}) as a function of the impact parameter 
$y$ and the redshifted lens mass $M_{\rm{Lz}}$.   
From top to bottom, the color bar represents the ratio of the lensed SNR $\rho_{\rm L}$ and unlensed SNR $\rho_{\rm UL}$, the signal-consistency test $\xi^{2}/\rho^{2}$ (see Eq.~\ref{search2}), and the FAR, respectively. 
An event is considered a significant gravitational-wave candidate when its FAR is below $3.85\times10^{-7}$ Hz or one every 30 days, indicated by the red tick.  
The source distances from left to right panels are $125$, $625$, and $1250$ Mpc. 
A large fraction of the lens parameter space does not pass the FAR threshold even at high SNR.}
 \end{figure*}

\section{Results}\label{Section: Results}
Our results highlight the impact of lensing on the matched-filter SNR and the $\xi^{2}$ value. 
This influence scales with lensing amplification and signal strength, an effect not accounted for in optimal SNR analyses. 
In the following sections, we explore these effects and provide a general overview of the detectability of lensed gravitational waves.

\subsection{The role of lensing}\label{sec:result_A}
We compare the recovered distributions of $\mathcal{M}_c$ and $\chi_{\rm{eff}}$ for sets $1$-$4$ in Fig.\ref{Fig: set1_recovered_template}. 
At 125 Mpc, unlensed injections have an expected SNR of approximately 100 at O3 sensitivity, with minimal noise affecting the recovered template parameters. 
The unlensed (blue) and set 2 (red) injections exhibit negligible differences, both centered at $\mathcal{M}_c = 26.7M{\odot}$ and $\chi_{\rm{eff}} = 0$. For $M_{\rm{Lz}} = 10^{3}M_{\odot}$ (green), $\chi_{\rm{eff}}$ shifts to approximately 0.25, while $\mathcal{M}_c$ remains qualitatively similar to unlensed injections. 
Set 4 (purple) shows significant lensing amplification, resulting in a substantial drop in recovered $\mathcal{M}c$ and a multimodal $\chi{\rm{eff}}$ distribution peaking at ${-1, 0, 1}$. 
This demonstrates that at $\rho \sim 100$, strong lensing distortion can significantly deviate recovered parameters from injected values, potentially reducing the matched-filter SNR. 
However, the right panel in Fig.\ref{match_rho} indicates an expected increase in SNR due to amplitude enhancement. 
Therefore, the resultant matched-filter SNR reflects a competition between the inaccuracy of recovered source parameters (reducing SNR) and lensing amplification (increasing SNR).

Figure~\ref{Fig: set1_chisq_snr_far} illustrates the competing effects of lensing on detectability by showing $\xi^{2}/\rho^{2}$ vs $\rho$ for sets $1$-$4$, with FAR values represented as the color map.
The unlensed scenario (upper left) demonstrates a reduced FAR with higher SNR and decreasing $\xi^{2}/\rho^{2}$, with 17020 out of 19432 injections passing the detection threshold of $\rm{FAR}_{\rm{thr}} = 3.85\times10^{-7}$ Hz.
Note that even in the representative unlensed $30$–$30$~$M_{\odot}$ spinless BBH at a fixed source distance, sampling over other source parameters and different noise realizations can result in a wide range of SNR values, subsequently leading to a wide range of $\rho$ values from $10$ to $100$.
Thus, this illustrative unlensed injection set already demonstrates that the actual matched-filter SNR differs from the optimal SNR.
Moreover, small lensing distortions in the upper right panel show a similar distribution with slightly more successful detections (17068), attributed to minor signal amplification and negligible waveform distortion.

The intermediate lensing distortion in the bottom left panel illustrates, for the first time, the competition between the inaccuracy of the recovered source parameters and lensing amplification.
Although we observe an overall increase in the SNR distribution due to lensing amplification, we also find a noticeable increase in the density of low-SNR injections with $\rho < 10$.
Thus, the resultant SNR, influenced by the above-mentioned competition, depends on other factors such as noise realizations and the source parameters sampled in the injection set.
After the SNR computation, the unlensed template used to retrieve the lensed injections constructs an expected SNR time series, while the actual SNR time series contains lensing distortion effects.
As a result, calculating $\xi^{2}$ using Eq.(~\ref{search2}) reflects this lensing distortion as an elevated $\xi^{2}$ value.
Although $\xi^{2}$ is designed to mitigate loud noise triggers, it does not distinguish whether deviations in the SNR time series arise from noise or waveform distortions caused by missing physics.
Consequently, we observe an overall increase in $\xi^{2}/\rho^{2}$, with the injections at higher SNR exhibiting the most significant enhancement, increasing from $\xi^{2}/\rho^{2} \sim 10^{-4}$ to $\xi^{2}/\rho^{2} \sim 10^{-3}$.
Since $\rho$ and $\xi^{2}$ play a key role in FAR computation, this increase leads to an overall enhancement in FAR, subsequently reducing the number of successful detections to 14251.

Our injection set with the strongest lensing distortion presents a scenario where only $1$ out of $19421$ injections is successfully detected.
A remarkable fraction of lensed injections is observed at $\rho_{\rm{opt}} < 10$ due to the poor accuracy of the recovered source parameters.
This inaccurately recovered template further constructs an SNR time series for $\xi^{2}$ computation, resulting in a $\xi^{2}$ value that is not designed to handle such large waveform distortion scenarios.
As a consequence, this leads to a significant increase in FAR, rising from $< 10^{-10}$ to $> 1$ Hz, and ultimately results in only one successful detection.

These results demonstrate that significant waveform distortion impacts the behavior of matched-filter SNR, suggesting that optimal SNR poorly represents detectability even at $\rho_{\rm{opt}} \sim 100$.
Furthermore, employing a large template bank with millions of templates in a realistic matched-filtering setup spans a broader source parameter space but only yields a wider recovered source parameter space.
This broader recovered space leads to a more inaccurate SNR time series for $\xi^{2}$.
Thus, our initial injection campaign results also support the use of a smaller template bank in detectability studies.

Fully understanding the inaccuracies in the recovered source parameters for the lensed injections, it is essential to conduct a detailed investigation of each template’s response. 
This includes analyzing metrics such as the SNR time series and the $\xi^2$ time series in response to the lensed injections. 
However, such an investigation is computationally intensive, requiring significant computational resources and time to process and analyze the data comprehensively. 
Consequently, this level of detail falls beyond the scope of the current study and is deferred to future work.

\subsection{The role of the signal strength}\label{sec:result_B}

Figure~\ref{Fig: set2_recovered_template} compares the recovered $\mathcal{M}_c$ and $\chi_{\rm{eff}}$ distributions for unlensed injections at 125 Mpc (blue) and lensed injections at 125 (purple), 625 (orange), and 1250 Mpc (green).
All lensed injection sets exhibit significant deviations from the unlensed distribution at all distances, demonstrating that substantial lensing distortion impacts the accuracy of the recovered source parameters against background noise.
However, the discrepancy between lensed and unlensed distributions decreases with increasing source distance, suggesting that matched-filtering searches are less sensitive to large waveform discrepancies in noisier backgrounds.
Consequently, we anticipate a smaller SNR loss than observed in the first injection campaign when the signal SNR is lower.

Figure~\ref{Fig: set2_chisq_snr_pass_lost} compares the detectability of unlensed injections (upper) and lensed injections with $y=0.01$ and $M_{\rm{Lz}}=10^{5}M_{\odot}$ (bottom) at distances of 125 (left), 625 (middle), and 1250 Mpc (right).
For the unlensed injections, larger source distances lead to smaller SNR and higher $\xi^{2}$ values.
Consequently, 17020, 12878, and 6217 unlensed injections were found for $d_{L}={125, 625, 1250}$~Mpc.
However, lensed injections exhibit an opposite trend with increasing distance.
As predicted in our analysis of the accuracy of recovered source parameters, more lensed injections are recovered with larger matched-filtering SNRs compared to the unlensed scenario when the signal strength decreases.
As a result, the effect of lensing distortion on SNR loss is mitigated in a noisier background.

Moreover, the $\xi^{2}$ value dominates at all distances, indicating that signal strength does not play an important role. 
High FAR values primarily result from large $\xi^{2}$ values caused by waveform distortion, regardless of signal strength.
Therefore, 1, 8, and 13 lensed injections were found for $d_{L}={125, 625, 1250}$~Mpc.
This injection campaign demonstrates that matched filtering more readily rejects signals that significantly deviate from the template at higher SNRs, potentially decreasing detectability even with improved detector sensitivity.

\subsection{General picture of detectability of lensed gravitational waves in matched-filtering searches}\label{sec:result_C}

To illustrate the general impact of lensing on detectability, Fig.~\ref{set3_all} displays the matched-filter SNR ratio of lensed to unlensed injections, the $\xi^{2}$ value, and the FAR within the $y$ vs. $M_{\rm{Lz}}$ lens parameter space.

Contrary to predictions from optimal SNR analysis that $\rho_{\rm{L}}/\rho_{\rm{UL}} > 1$, the top upper panel of Fig.~\ref{set3_all} reveals a purple region where $\rho_{\rm{L}}/\rho_{\rm{UL}} < 1$.
As shown in the first injection campaign, there is a competing effect between the inaccuracy of the recovered source parameters caused by waveform distortion (reducing SNR) and lensing amplification (enhancing SNR).
Therefore, this purple area underscores a region where the inaccuracy of the recovered source parameters dominates over the lensing amplification.
Notably, this region aligns with $\mathcal{M} \lesssim 0.75$ on the left panel of Fig.~\ref{match_rho}, further suggesting a possible correlation between matches and the accuracy of the recovered source parameters, which will be explored in future studies.

Outside the purple region, the SNR ratio of lensed to unlensed injections aligns quantitatively with optimal SNR analysis.
The right panel of Fig.~\ref{match_rho} indicates an area where lensing amplification dominates over the inaccuracy of the recovered source parameters.
In addition, the $\rho_{\rm{L}}/\rho_{\rm{UL}} < 1$ region shrinks with increasing source distances when we compare the SNR ratios from the top left to the top right panel.
This observation aligns with the results of the second injection campaign, where the impact of lensing distortion in the high-SNR regime ($\rho_{\rm{opt}} \sim 100$) is the primary contributor to the inaccuracy of the recovered source parameters.
This contribution from lensing distortion is mitigated by background noise when the signal strength is less prominent ($\rho_{\rm{opt}} \sim 10$). 
 
The middle left panel of Fig.~\ref{set3_all} illustrates how different lensing parameters influence the $\xi^{2}$ values at $125$~Mpc.
We observe that the region with substantial lensing distortion leads to high $\xi^{2}$ values, which also aligns with the region where $\rho_{\rm{L}}/\rho_{\rm{UL}} < 1$ in the top panel of the same figure.
This suggests a region of extensive increases in FAR within similar regions of the lens parameter space.
Furthermore, we observe an expected increase in $\xi^{2}$ at larger source distances due to enhanced background noise, indicating how background noise can mitigate the impact of waveform distortion in matched-filtering searches.

The bottom left panel of Fig.~\ref{set3_all} confirms our expected region with a large FAR, aligning with the SNR loss and high $\xi^{2}$ regions in the upper and middle panels.
This demonstrates that the SNR loss and enhanced $\xi^{2}$ caused by strong lensing distortion collectively contribute to the low detectability of lensed gravitational waves.
From the bottom left to the bottom right panel, there is an overall increase in the number of lensed injections with FAR values larger than the detection threshold.
This is a direct consequence of the impact of weaker signal strength, where stronger background noise results in less significant ranking statistics when SNR, $\xi^{2}$, and all other parameters in Eq.(~\ref{likelihood_ratio_ranking}) are considered.
In addition, we observe that the high FAR region decreases with increasing source distances from the bottom left to the bottom right panels, which aligns well with the results in the upper and middle panels.

The third injection campaign yields crucial insights.
First, optimal SNR analysis fails to account for the substantial SNR loss caused by lensing distortion and the incorporation of signal-consistency tests.
This information is critical for assessing the detectability of lensed gravitational waves in wave optics regimes. 
Lensing analyses and lensing rate calculations, which are based on the optimal SNR method, overestimate detectability in the wave optics regime.
Moreover, despite the prominent drop in detectability, lensed signals that significantly deviate from the unlensed template due to wave optics effects are more detectable with weaker signal strength in matched-filtering searches.
Consequently, improving detector sensitivity may not enhance the detectability of lensed gravitational waves.
Figure~\ref{set3_all} also highlights the need for a lensed gravitational-wave template bank, offering valuable guidance on optimal lensing parameter space regions for future template development, particularly for the point mass lens model.

\section{Conclusion}\label{Section: Conclusion}
 
The detectability of new types of gravitational-wave signals with matched-filtering searches depends on a complex interplay between the completeness of the template bank and the background noise.

A reduction of the detectability can be prominent when the  
missing physics in the original template bank introduces significant waveform distortions to the gravitational-wave signal, such as beating patterns in gravitational lensing in the wave-dominated regime~\cite{Takahashi:2003ix}.
One conventional approach to approximate the detectability of such signals is to compute the optimal SNR. However, this statistic does not account for the possibility of the noise mimicking a real signal. 
Other detection statistics, such as the signal-consistency test value, are precisely included in the search pipelines to mitigate these effects. 
A comprehensive study of the background noise in the detectors is also necessary to obtain the probability that a given signal is real. 

For the first time, we study the detectability of lensed signals with a state-of-the-art matched-filtering search pipeline, capturing the detailed information in the detection statistics and real background noise effects.
We ``inject'' simulated lensed gravitational-wave events into data with real noises and retrieve them using a time-domain matched-filtering search pipeline, GstLAL~\cite{2017PhRvD..95d2001M,Sachdev:2019vvd,2021SoftX..1400680C,Tsukada:2023edh}. 

We perform three distinct injection campaigns to determine the detectability of lensed gravitational waves.
The first two injection campaigns study, respectively, the role of lensing and the role of the signal strength, while the third one investigates the general picture of the detectability of lensed gravitational waves. 
We observed that matched-filtering searches using the unlensed templates could retrieve a lensed and magnified signal with a higher SNR in the first two injection campaigns, which, for a fixed impact parameter, encompass different lens masses and source distances.
However, a fraction of lensed signals loses SNR due to inaccurate recovered source parameters in the search pipeline.
The number of signals with a decreased SNR scales with both the lensing amplification and the signal strength.
One representative scenario is a significant fraction of highly magnified and distorted lensed gravitational waves ($y=0.01$, $M_{\rm{Lz}}=10^{5}M_{\odot}$) at $125$~Mpc (with an unlensed SNR $\sim 100$) being retrieved with matched-filter SNR $<10$.

In addition, the autocorrelation-based signal-consistency test value $\xi^{2}$ plays a crucial role.
Searching for lensed gravitational waves with an unlensed template bank can increase the $\xi^{2}$ value to dominate the detectability due to substantial waveform distortion. 
With the loss in SNR and increase in the $\xi^{2}$ value, which can increase the false alarm probability by many orders of magnitude, the detection efficiency of a signal at 125~Mpc drops from $\sim90\%$ (unlensed) to $< 1\%$ (with $y=0.01$, $M_{\rm{Lz}}=10^{5}M_{\odot}$). 
These two effects, which optimal SNR analyses do not predict, suggest that lensed gravitational waves are less detectable in matched-filtering searches.

As shown in the third injection campaign in Sec. \ref{sec:result_C}, lensed gravitational waves in the wave optics regime, which introduces the most significant waveform deviation, deviate from optimal SNR analysis.
This highlighted that the wave optics regime has the most significant impact on gravitational-wave signals, guiding future efforts in lensed template bank construction. 

Based on the insights gained from our injection campaigns, we suggest the following strategies for detecting the lensed gravitational waves. 
First, matched-filtering searches with a lensed template bank are necessary to probe the lens signatures, especially in the regions of large waveform distortions.
Construction of such a template bank requires a new injection campaign that further samples all source parameters, as well as a critical assessment of the computational efficiency of extending the parameter space of the template bank.  
Second, we suggest that coherent wave burst pipelines can offer an alternative path in detecting lensed gravitational-wave signals due to its model-independent nature~\cite{Drago:2020kic}.
Unlike matched-filtering searches, coherent wave bursts can identify loud signals without relying on predefined waveform templates, making it more likely to capture lensing-induced modifications.
Performing follow-up lensing analyses on gravitational-wave candidates found only by coherent wave burst pipeline presents an avenue for detecting lensing signatures that may have been overlooked by template-based searches.
This approach complements matched-filtering techniques and could provide a more comprehensive view of lensed gravitational-wave events.

Since the detectability of lensed gravitational waves in matched-filtering searches differs from optimal SNR analyses, there are several consequences for other lensing analyses.  
First and foremost, the detectability of lensed gravitational waves in the wave-dominated regime is significantly lower in the unlensed matched-filtering searches compared to the optimal SNR approximation. 
Any follow-up parameter estimation that probes the wave optics signatures can suffer from significant bias, as they are only looking at signals already in current catalogs, and these will not include highly distorted lensed signals. 
Properly accounting for these selection biases requires a large-scale injection campaign that further samples all the source and lens parameters to specify the probability of detection in the parameter estimation.

Additionally, reassessing the lensing detection forecasts and constraints based on optimal SNR is necessary.
Current lensing rate estimations are based on optimal SNR detection thresholds~\cite{Meena:2022unp,Mishra:2021xzz,Meena:2022unp,Mishra:2023ddt}.
Although significant lensing distortions are rare, current analyses overestimate the detection rate of lensed gravitational waves.
Furthermore, the same applies when setting constraints on the source and lens populations from the (non)observation of lensing in the gravitational-wave catalogs. 
This is particularly relevant for compact dark matter scenarios, which could constitute an effective population of lenses for gravitational waves in the wave-dominated regime. 
In the past, constraints on those scenarios have been performed based on optimal SNR analyses~\cite{Jung:2017flg,Diego:2019rzc,Liao:2020hnx,Oguri:2020ldf,Urrutia:2021qak,Tambalo:2022wlm,Zhou:2022yeo}.
Since these constraints do not correctly include the reduced detectability in matched-filtering searches shown in our work, the current constraints are overoptimistic in some parts of the parameter space. 
A full injection campaign is required for these reanalysis.

Searches of new physics in gravitational-wave observations may require dedicated pipelines.
This work showcases a scenario in which traditional optimal SNR analysis deviates significantly from matched-filter searches when the new physics are not included in the template.
Focusing on the case of gravitational-wave lensing, we also show how missing physics impacts the detectability of recovered source parameters and signal-consistency tests.
Our methodology is, however, more general and can be applied to other problems of interest.
In the future, we will explore this further, performing other injection campaigns to follow up on how different physics, such as deviation from general relativity, impact the detectability.
Such investigation can unveil the complex dynamics of how noises and different physics correlate in matched-filtering searches and, hence, help improve the corresponding detection method. 

\begin{acknowledgements}
The authors would also like to acknowledge Kipp Cannon, Luka Vujeva, Rico Ka Lok Lo and Elwin Ka Yau Li for their useful suggestions.
J.C.L.C. and J.M.E. acknowledge support from the Villum Investigator program supported by the VILLUM Foundation (grant no. VIL37766 and no.~VIL53101) and the DNRF Chair program (grant no. DNRF162) by the Danish National Research Foundation. 
E. S. is supported by grants from the College of Science and Engineering of the University of Glasgow. 
K.Y.A.L. would like to acknowledge the support from Croucher Foundation. 
H.F. acknowledges support from the CITA National Fellows program, the NSERC DG program, and the Canada Research Chairs program.
This project has received funding from the European Union's Horizon 2020 research and innovation programme under the Marie Sklodowska-Curie grant agreement No 101131233. 
J.M.E. is also supported by the Marie Sklodowska-Curie grant agreement No.~847523 INTERACTIONS. 
The Tycho supercomputer hosted at the SCIENCE HPC center at the University of Copenhagen was used for supporting this work. 
This research has made use of data or software obtained from the Gravitational Wave Open Science Center (gwosc.org), a service of the LIGO Scientific Collaboration, the Virgo Collaboration, and KAGRA. This material is based upon work supported by NSF's LIGO Laboratory which is a major facility fully funded by the National Science Foundation, as well as the Science and Technology Facilities Council (STFC) of the United Kingdom, the Max-Planck-Society (MPS), and the State of Niedersachsen/Germany for support of the construction of Advanced LIGO and construction and operation of the GEO600 detector. Additional support for Advanced LIGO was provided by the Australian Research Council. Virgo is funded, through the European Gravitational Observatory (EGO), by the French Centre National de Recherche Scientifique (CNRS), the Italian Istituto Nazionale di Fisica Nucleare (INFN) and the Dutch Nikhef, with contributions by institutions from Belgium, Germany, Greece, Hungary, Ireland, Japan, Monaco, Poland, Portugal, Spain. KAGRA is supported by Ministry of Education, Culture, Sports, Science and Technology (MEXT), Japan Society for the Promotion of Science (JSPS) in Japan; National Research Foundation (NRF) and Ministry of Science and ICT (MSIT) in Korea; Academia Sinica (AS) and National Science and Technology Council (NSTC) in Taiwan.
\end{acknowledgements}

\bibliographystyle{apsrev4-1}
\bibliography{citations}
\end{document}